# How to give a great talk

Chat Hull

NAOJ Fellow, National Astronomical Observatory of Japan, Joint ALMA Observatory

The art of the scientific presentation—much like the art of the perfect plot, the art of the compelling proposal, and the art of the killer job application—is generally not something we're taught in school. So, in classic Millennial style, one of the first things I did when I was asked to write this piece was to ask Google, "how to give a scientific presentation" to see what the Internet had to say. I've attempted to avoid boring you with laundry lists of "do's" and "don'ts," as many of the top Google hits did [1], but I do incorporate a wide range of tips—and some interesting vignettes—both from the online hive mind and from my ~8 years of regularly giving astronomy and radio science presentations.

**Practice and timing**

We all know the adage that 10,000 hours of practice makes you an expert. While 10,000 hours of rehearsal for a single colloquium is probably overkill, a few hours go a long way. In short, *practice is critical!* I've gotten pretty good at giving talks after nearly a decade of giving over 100 presentations, but some people are still surprised to hear that before every presentation I give, I practice at least once, if not two or three times, depending on the importance (i.e., job interview versus 15-minute lunch seminar) and the amount of new material I'm covering.

Practice is a magical thing: it makes you more familiar with your own material; it gives you a chance to explain concepts out loud instead of in a questionably formatted LaTeX document; it allows you to smooth out the flow and arc of your story telling; and it allows you to calibrate your timing. Speaking of that: in this era of infinite distraction and limited attention span, this is arguably one of the most important pieces of advice I can give: *don't exceed your allotted time!*

For conference talks (e.g., 12 minutes + 3 minutes for questions) I plan to stop right at 12 minutes (much to the delight of the session chair), since audiences tend not to ask questions until the end. However, for a 60-minute colloquium I'm much more generous, usually planning only about 40 minutes of material without questions. (Note that I don't abide by the 1-slide-per-minute rule, partially because many of my animations, zoom-ins, and overlays tend to be composed of series of slides.) It's a win-win: if the audience peppers you with questions, you have plenty of time to answer all of them and still end within the 60 minutes. And if the audience is full of question-less zombies, then you end 15 minutes early: nothing wrong with that!

All of us have secretly cringed when the opposite happens: the speaker plans unwisely for 60 minutes of material, but after fielding a few on-the-fly questions is on track to go 10 minutes over. Toward the end, if the chair is overly permissive, the speaker simply plows forward, and the audience slowly starts to dissipate through the rear exit. However, if the chair is stricter, the outcome is arguably even worse: the speaker arrives at the conclusions—the most important section of the talk—*and completely glosses over them!* Not wise in the ways of science. Note that even if you do go over time, you can mitigate this problem by stating your conclusions at the very beginning of the talk (and in the middle…and as many other times as possible, because people *will* periodically fall asleep!). Don't be one of those speakers who says, "I'm withholding the conclusion until the end so that you'll pay attention": that's a recipe for disaster.

Strategic note: one way to maximize the likelihood of timing nirvana is to check in advance with your host and find out how interactive the audience tends to be (and, even better, who the usual question-askers are); that can help you adjust both your timing and your preparation of backup slides that address detailed questions that aren't worth covering in your talk.

Here's my general recipe for practicing talks. Once you've got a decent draft up and running, go through the talk for the first time. Don't time it. As you're making your way through it, correct small text errors, slide order, animation order, and other things that interrupt your flow. Note: especially if you've added a lot of new material or are presenting work that's less familiar to you, *this first run-through will be terrible!* Don't despair. Press on: go get a cup of coffee, and once you've recovered from feeling like you don't understand anything about the paper you just wrote, stand up and go through the presentation again. You'll be astounded by how much better it is the second time. This time, time it. If the timing is about right, then you're probably good to go. If you go way over, go back and delete some slides (or hide them, or move them to the "extra slides" section after the talk) and run through it one more time. A final note: even if you have a polished talk that you're going to give several times in quick succession, try to practice at least once within 24 hours of the talk (ideally right before you go to bed the night before); then, within a few hours of the talk make sure to find 3 minutes to sit down and click through all of your slides without speaking, just to pre-load the information into your brain.

At this point, you're totally ready…so ready that you can actually punctuate your slide transitions with words that emphasize the material you're showing. Not only will knowing your slide order make you feel more at ease, but it will impress the audience. When I'm watching a speaker who obviously doesn't know which slide is coming up, I tend to tune out because it's obvious that they didn't care enough to prepare. It's just like being a journal referee and getting a draft full of typos that a spell-checker could have caught. Don't waste the audience's time.

Under some circumstances—like faculty interviews—it pays to actually memorize your slide numbers (or at least have a note card on hand with the few most important ones). In both PowerPoint and Keynote you can type a number and then hit Enter, and it will take you right to that slide. The new MacBook Pro's tiny "Touch Bar" makes this even easier. All of these tips will save you from the inglorious few seconds of revealing your desktop—or worse, clicking through 85 animations—while you search for a particular slide.

**Knowing your audience and tweaking your delivery**

I did a grand job-talk tour when I was applying for my first postdoc, and before I arrived at each institution I scanned the faculty page to see who was interested in star formation. Most of them I would email in advance—*especially* if they were known for having a penchant for shredding speakers whose ideas conflict with their own personal views of the field. This has worked wonders for me any number of times: at one point during my tour I invited a particularly notorious professor to my talk, at the end of which they merely asked me one or two productive, positively-spun questions. My host was astounded, not only because the professor made a rare appearance at a seminar, but because they also seemed to have a genuinely positive reaction to my work. Another less involved version of this tip is: before you give a talk (and *especially* before a job interview), look over a few of the most important publications from each faculty member who is likely to be interested in your work…and then weave in appropriate verbal (or even written) citations to the work in your talk. Never underestimate the power of showing people that their work, presence, experience, and professional opinions are appreciated.

In addition to that, never underestimate the power of telling people things they already know (science is hard—it's good to feel smart every now and then!). I was recently invited to give a colloquium, and the invitation specifically stated:

> *We encourage you to present a general colloquium with ample introduction to the topic, rather than a specialized talk that would appeal only to experts. Don't underestimate the pleasure physicists and astronomers take in hearing about subjects that are already familiar to them.*

People have varying opinions about the ideal breakdown of a talk in terms of expertise level, but I generally try to adhere to the rule of thirds: the first third of the talk should be extremely general (in my case, "stars form out of blobs of dust and gas"), 100% understandable by everyone in the audience. The second 33% can be aimed at experts in star formation, but not necessarily those who work very closely with you. Then the last bit of the talk might be aimed at the few people who you've come to collaborate with. Although, in my opinion, it's best if you avoid focusing that sharply at the end, aiming instead to keep as much of the audience as possible engaged for the entire time.

As for actually *giving* your talk, the first vague piece of advice I usually give is, "be yourself!" I'm a total ham, so I make jokes. But if you're not a ham, then don't! It's okay: crafting a presentation style consistent with your own personality is far easier than forcing yourself to be someone you're not.

In that vein, however, one area where I truly believe many scientists *should* sometimes deviate from their own typical behavior is their appearance: *dress up!* Show the audience that you care! I admit that I was educated for ten years in the state of Virginia, where I wore a coat and tie at least once a week (and a suit or a tux at least once every couple of months). However, I believe that even if you're a wrinkly-T-shirt-and-shorts kind of person, as many scientists are, that doesn't mean that you have to look wrinkly for your talk. Much as knowing your slide transitions shows that you care about the audience's time, wearing decent clothes (even just an ironed shirt and pants/skirt with a pair of non-sneaker shoes) shows that you respect them enough to professionalize your appearance for the hour that they're dedicating to your presentation.

Regarding presentation style, there is an art to striking the right tone. The first presentation I gave in graduate school was in the much-feared Journal Club seminar, where you have to go through the somewhat unnatural process of presenting a summary of a recently published paper as if you had written it yourself. As a first-year graduate student who didn't even know what an early-type galaxy was, this was quite difficult, and after I finished my presentation, the faculty member in charge of Journal Club came up to me and said, *"you have to stop saying, 'I don't know'!"* That feedback has stuck with me and forever changed how I set the tone of my talks. I think the faculty member was wrong to tell me that in such a general way; however, there is some truth to the sentiment, because we as scientists need to strike the right balance between being totally unsure of our work and being so confident that we dismiss constructive comments from our colleagues. My way of dealing with this is generally to request questions before (and sometimes during) the talk, so that as the talk progresses the audience gets comfortable bantering with me about the material. Even more important, I truly believe it is appropriate to admit ignorance under the right circumstances. Adding, "I don't know" or "that's a great avenue to pursue, let's talk afterward" to a thoughtful but inconclusive response to a question is a good way to show that you know what you're talking about, but you don't know everything, and you're willing to consider (and you appreciate) the intelligent ideas of your fellow scientists.

A few other things I focus on stray into the realm of oratory technique. First: look people in the eye when you speak; it engages them, and makes it less likely that you'll stand with your back to the audience reading the slides. And second: *slow down*. One of my favorite parts of astronomy is its global nature: we come from everywhere, and as a result, a vast number of astronomers speak English as a second (or third, fourth, fifth…) language. Thus, it pays to speak slowly; furthermore, while fancy vocabulary can be fun, it should be used sparingly so that as many readers and listeners as possible can understand our work. In addition to speaking slowly, a well-placed pause can be a nice reprieve for the audience after several dense slides. Occasional silence is okay: it gives people a few moments to absorb what you just said (or the plot that you have strategically left up on the projector); and it gives you a second to take a drink of water!

**Crafting a great presentation**

Now we come to the technical details of how to make an excellent presentation. In some regards, this is the area where opinion matters least (in my opinion[!]), because there are cognitive research studies showing that certain color schemes and slide layouts that are objectively better than others. There is no shortage of advice out there, and I recommend that you peruse it at your leisure, from Edward Tufte's books on data visualization, to websites like Presentation Zen [2], to presentation style recommendations from the creators of PowerPoint themselves [3].

I personally subscribe to the "less is more" philosophy, and use as little text as possible (approaching, if not quite achieving, a terse "Lessig-style" presentation [4] with only a few words per slide). A sage colleague of mine once said, "pixels are finite; slides are infinite." In short: make your images HUGE, annotating them with just the little bit of text you need to make them understandable. That way the audience will listen more closely to your interpretation of the plot (which they'll be able to see because it's so huge), and you won't be tempted to stare at the screen and repeat the words on the slide (which the audience will have finished reading long before you've finished saying them).

**Avoiding technical problems**

It is well known that multivariate Calculus is far easier for most scientists than using an overhead projector. Invariably, this leads to a substantial amount of lost time at every conference, and at a non-negligible fraction of colloquia and seminars. While it is impossible to prepare for all technical glitches, there are a few things you can do that will help you avoid a complete technological meltdown and will make your presentation flow much more smoothly.

The first thing to do is to figure out what the projection situation will be during your presentation. If you're presenting at a conference, the website will usually have instructions regarding whether transferring your presentation file is necessary and what sort of laptops (e.g., Mac, PC) and programs (e.g., Keynote, PowerPoint) will be available. If you can't easily locate this information (or if you're giving an informal lunch talk somewhere), then you should send an email to inquire well in advance so you can prepare. In most cases, if you have an up-to-date version of Keynote or PowerPoint, all will be well, especially if you can use your own laptop. When I used PowerPoint on my MacBook Pro I virtually never had any issues. The only issue I've had since switching to Keynote about a year ago was one instance where I was required to present using PowerPoint on a conference-supplied machine, and the conversion from Keynote to PowerPoint mangled some of the fonts I was using. A way around that is to use extremely basic fonts; however, in my experience it's very rare to be forced to convert your presentation file. Note that my generally-problem-free experience has been enabled by being a stereotypical scientist who uses an expensive MacBook with expensive Microsoft and Apple software: users of Linux machines and software such as OpenOffice should proceed at their own risk, but should also know that their pursuit of excellence with different software and machinery is appreciated by the open-access community!

My first order of business on the day of my presentation is to load my laptop bag full of all of the possible laptop accoutrements I might need, including all possible VGA, DVI, and HDMI dongles; a USB stick for transferring talk files; and my own clicker/laser pointer (I recommend the Logitech Wireless Presenter with a *green laser pointer*—see below for a discussion of laser color). Having multiple dongles can solve projector issues (e.g., if for some reason the projector works with a VGA connection but not with HDMI), and also solves the problem created by Apple's constant modification of their laptop ports (and the fact that, for example, some older MacBook Air models have no HDMI port, whereas the same generation of MacBook Pros do). For talk flow, I find that having my own pointer/clicker is critical, mainly because I know where all the buttons are, so I can easily use the laser pointer and click through my slides without accidentally blanking the screen or mashing a button that causes my presentation to skip forward mysteriously by 13 slides.

Another good thing to do before plugging in the projector is to turn off your wireless connection, quit other running programs, and disable all desktop notifications: this will help avoid both annoyance (when your computer repeatedly complains about not being able to connect to the guest wireless network) and embarrassment (when your calendar automatically reminds you about your upcoming dentist appointment…or perhaps something less benign).

When the moment of truth comes and you plug in the screen, 80% of the time it will work. For the other 20%, it really, really pays to know how your laptop works (easier said than done, I know)…again: practice, practice, practice! Over the years, I've gotten very good at manipulating the Display preferences under "System Preferences" on my Mac, tweaking everything from display mirroring to output resolution to the physical scaling of the image so that the bottom of my slide doesn't get cut off. Know your machine.

And finally, if all hell truly breaks loose, you can always resort to exporting your presentation to a PDF and presenting it on someone else's computer (remember your handy USB stick!). Just remember to avoid overlapping animations: by splitting progressive animations into multiple slides you'll be able to simulate the animation even in PDF mode and you won't confuse audiences with a single slide featuring 14 randomly overlapping plots that faded in and out in your original PowerPoint file. Of course, in a PDF you won't be able to show movies or have your text fly in dramatically from overhead, but you'll still be able to transmit the crux of your scientific story, which is all that matters.

**A few more suggestions**

Now I will indulge myself and outline a few remaining items on my wish list for great talks:

Equations: don't show them. Maybe $\theta \sim \lambda/D$ is okay, but nothing more complicated. Just my seasoned observer's opinion.

Citations: cite everything—every image, plot, photo, etc. Many of your talk files will end up on the Internet…the more attribution the better.

Animations: use them to your advantage, and use them to guide the eye. If you must use substantial amounts of text (e.g., in the conclusion slide—which, by the way, should still feature representative plots as visual summaries!), then animate each bullet independently so that the audience doesn't get distracted by the rest of the text.

Be colorblind friendly: this is my personal soapbox, as I suffer from mild colorblindness along with ~10% of the white male population. First: use a green laser pointer; red ones are invisible (though I admit that it's better to use green lasers that have very low wattage, as the normal stargazing pointers can be exceptionally bright and distracting). Second: when you make plots, use bright colors—red and black dots are indistinguishable, for example. However, if you must use similar colors, use different symbol styles (e.g., red triangles and black squares). Third: use highly contrasting text with respect to your slides' background color. Red on black = bad. Blue on black = bad. Yellow on white = super-bad. I don't know how many times I've seen people try to highlight the critical word in a sentence in red, which was thus been rendered unreadable on the black background! When in doubt, consult the outstanding Colorbrewer website [5], which brews up the perfect, colorblind-safe color combinations for your plots, complete with hex codes!

And finally, regarding plot complexity: when it comes to showing plots in a talk (versus in a paper), I abide by the KISS method ("Keep it simple, stupid!"). Don't dwell on technicalities: if a detail of a plot is likely to be distracting or is irrelevant, either cover it up, crop it out, or make a new plot for your talk. Also, axis labels on plots copied from papers are almost always too small: crop the original axes and replace them with HUGE text with something intuitive like "Brightness," as opposed to, say, $erg/s/cm^2/Hz$.

Note that the above advice is relevant mainly for the default presentation style today, in the late 2010s: Keynote or PowerPoint slides on a big projector. But of course, many other presentation styles exist: shared screens on video-cons and Skype interviews; chalk talks; posters at conferences; etc. For example, when you're giving an online interview, you'll want to make sure to build in time for the interviewers to flip through your slides, and you'll want to practice saying, "next slide…next slide…" and, "on the bottom of slide 13 you see…"

**A few final words**

And that, my fellow scientists, is my current understanding of how to give a great talk. I hope the above stories and tips form a coherent picture of how to craft understandable slides, practice effectively, learn the ins and outs of your audience, avoid technological catastrophe, and execute your talk on time and in an engaging manner. At some level, nearly all of this is subjective, and the breakneck pace of technological development means that we'll all need to revisit this topic regularly throughout our careers. But for now, onward and upward, practice-practice-practice, and break a leg!

*About the author*

*Chat Hull has been giving scientific presentations of various sorts since his days as an undergraduate at the University of Virginia, and later as a summer research student at the University of Rochester, a high school professor in both Virginia and Guatemala, an astronomy graduate student at the University of California, Berkeley, and a Jansky Fellow of the National Radio Astronomy Observatory based at the Harvard-Smithsonian Center for Astrophysics. He is currently working for the National Astronomical Observatory of Japan as an NAOJ Fellow based at the Joint ALMA Observatory in Santiago, Chile.*